\documentclass[twocolumn,showpacs,preprintnumbers,amsmath,amssymb]{revtex4}
\usepackage{graphicx}
\newcommand{\ltsim}{\protect\raisebox{-0.5ex}{$\:\stackrel{\textstyle <}{\sim}\:$}}
\newcommand{\gtsim}{\protect\raisebox{-0.5ex}{$\:\stackrel{\textstyle >}{\sim}\:$}}
\begin{document}
\title{Photoinduced charge and spin dynamics in strongly correlated electron systems}
\author{H. Matsueda}
\author{S. Ishihara}
\affiliation{Department of Physics, Tohoku University, Sendai 980-8578, Japan}
\date{\today}
\begin{abstract}
Motivated by photoinduced phase transition in manganese oxides, charge and spin dynamics induced by photoirradiation are examined. We calculate the transient optical absorption spectra of the extended double-exchange model by the density matrix renormalization group (DMRG) method. A charge-ordered insulating (COI) state becomes metallic just after photoirradiation, and the system tends to recover the initial COI state. The recovery is accompanied with remarkable suppression of an antiferromagnetic correlation in the COI state. The DMRG results are consistent with recent pump-probe spectroscopy data.
\end{abstract}
\pacs{78.20.Bh, 78.47.+p, 75.47.Lx}
\maketitle

Competition among multiple phases is a key issue to understand electronic states in strongly correlated electron systems. Nature of the phases is characterized by internal degrees of freedom of electrons, namely spin, charge, and orbital~\cite{Maekawa}. Low-lying fluctuations of these degrees of freedom grow up toward the phase boundary. Then, tiny amounts of external perturbations can dramatically change the electronic states. Prototypical examples are perovskite manganese oxides where a ferromagnetic metallic (FM) phase competes with a charge-ordered insulating (COI) phase associated with antiferromagnetic (AF) and orbital orders~\cite{Imada}. In the COI phase, electric resistivity drastically decreases by magnetic field, pressure, impurity, and so on.

Photoirradiation is known to be a powerful tool to induce such dramatic response. This is called photoinduced phase transition (PIPT)~\cite{Tokura}. Ultrafast manipulation of femtosecond pulse laser is characteristic of the PIPT, and provides a striking difference between the PIPT and the other phase transitions. At early stages of the PIPT, only electronic part of the systems is strongly affected, and the spin, charge, and orbital degrees of freedom play major roles on relaxation dynamics. This is in contrast to the other phase transitions where structual deformation sometimes occurs. Thus, we expect novel relaxation processes of the PIPT and electronic structures that can not be created in equillibrium conditions.

The manganese oxides provide good playgrounds for elucidating effects of the multiple degrees of freedom on the PIPT~\cite{Miyano,Fiebig,Takubo,Ogasawara,Miyasaka,Liu}. In particular, the photoirradiation into the COI phase strongly influences transport, magnetic, and optical properties. Remarkable decrease of electric resistivity has been observed in Pr${}_{0.7}$Ca${}_{0.3}$MnO${}_{3}$ after the photoirradiation~\cite{Miyano,Fiebig,Takubo}. Metallic behavior has been also observed in pump-probe reflection spectroscopy measurements on Nd${}_{0.5}$Ca${}_{0.5}$MnO${}_{3}$~\cite{Ogasawara}: spectral weight of optical conductivity above the COI gap is transfered into lower-energy region within 400 fs after the photoirradiation. Magneto-optical Kerr spectroscopy measurements performed on Nd${}_{0.5}$Sr${}_{0.5}$MnO${}_{3}$ and Gd${}_{0.55}$Sr${}_{0.45}$MnO${}_{3}$ suggest that a FM component is induced by the photoirradiation~\cite{Miyasaka,Matsubara}.

The purpose of this Letter is to examine microscopic processes of the PIPT in a system where the charge and spin degrees of freedom strongly couple with each other. Motivated by the spectroscopy measurements on the manganese oxides, we examine effects of the photoirradiation on the COI  phase by calculating the transient optical absorption spectra in the extended double-exchange (DE) model. Theoretical treatment of many-body interactions, dynamical response, and time dependence due to nonequilibrium states on an equal footing is challenging~\cite{Takahashi,Maeshima}. In order to overcome this difficulty, we perform a density matrix renormalization group (DMRG) calculation~\cite{White,Hallberg,Kuhner,Jeckelmann,Matsueda,Cazalilla,Feiguin}. We find that the spectral weight above the COI gap is transferred into lower-energy region just after the photoirradiation, and then tends to go back to higher-energy region. This spectral-weight transfer is accompanied by remarkable suppression of AF correlations. In order to understand roles of localized spins on the transient spectra, the spectra of the extended Hubbard model are also examined. The DMRG results are consistent with the recent pump-probe spectroscopy data.

The Hamiltonian of the extended DE model is defined by
\begin{eqnarray}
H&=&-t\sum_{\left<ij\right>,\sigma}c^{\dagger}_{i,\sigma}c_{j,\sigma}+U\sum_{i}n_{i,\uparrow}n_{i,\downarrow}+V\sum_{\left<ij\right>}n_{i}n_{j} \nonumber \\
&&-J_{H}\sum_{i,s,s^{\prime}}\vec{S}_{i}\cdot c_{i,s}^{\dagger}(\vec{\sigma})_{ss^{\prime}}c_{i,s^{\prime}}+J_{s}\sum_{\left<ij\right>}\vec{S}_{i}\cdot\vec{S}_{j},
\end{eqnarray}
where $c_{i,\sigma}^{\dagger}$ ($c_{i,\sigma}$) and $\vec{S}_{i}$ represent a creation (annihilation) operator of a conduction electron with spin $\sigma$ at site $i$, and a localized spin operator, respectively. The electron hopping, the on-site and nearest-neighbor Coulomb repulsions are taken to be $t$, $U$, and $V$, respectively. The AF exchange coupling betwwen the nearest-neighbor spins is taken to be $J_{s}$ ($>0$). The conduction electrons interact with the localized spins by the Hund coupling $J_{H}$ ($>0$). We define $n_{i}=n_{i,\uparrow}+n_{i,\downarrow}$, $n_{i,\sigma}=c_{i,\sigma}^{\dagger}c_{i,\sigma}$, and $\vec{S}_{\rm tot}=\sum_{i}\vec{S}_{i}+\sum_{i,s.s^{\prime}}\frac{1}{2}c_{i,s}^{\dagger}(\vec{\sigma})_{ss^{\prime}}c_{i,s^{\prime}}$ with the Pauli matrices $\vec{\sigma}$. The magnitude of the localized spin is assumed to be $1/2$. From now on, the one-dimensional (1D) version of the model is taken into account. This simplification is reasonable, since a ground-state phase diagram does not depend so much on dimensionality~\cite{Yunoki}. Effects of the spin-charge separation on the photoinduced dynamics will be discussed later.

Let us define the transient optical absorption spectrum. First, pump light with energy $\omega_{0}$ comes into the system at time $t_{0}$. At $t_{1} (=t_{0}+\tau>t_{0})$, a transient state is described by
\begin{eqnarray}
\left|\psi(\tau)\right>&=&Ae^{-iH\tau}{\rm Im}\left(\frac{1}{\omega_{0}+E_{0}-H+i\gamma_{0}}\right)j^{\dagger}\left|0\right>,
\end{eqnarray}
where $\left|0\right>$ represents the ground state with energy $E_{0}$, $j=it\sum_{i,\sigma}(c_{i,\sigma}^{\dagger}c_{i+1,\sigma}-{\rm H.c.})$, $\gamma_{0}^{-1}$ corresponds to the dulation time of the pump light, and $A$ is a normalization factor. Next, the transient state is probed by a photon with energy $\omega$. The transient spectrum is given by
\begin{eqnarray}
\alpha^{\prime}(\omega,\tau)&=&-\frac{1}{\pi}{\rm Im}\left<\psi(\tau)\left|j\frac{1}{\omega+E_{1}-H+i\gamma}j^{\dagger}\right|\psi(\tau)\right>, \nonumber \\
\end{eqnarray}
where $E_{1}=\left<\psi(\tau)\right| H\left|\psi(\tau)\right>$ and $\gamma^{-1}$ corresponds to the dulation time of the probe photon. This form is valid for $\tau\gtsim\gamma_{0}^{-1}+\gamma^{-1}$. In order to illustrate the effect of the photoirradiation on the optical spectra, $\alpha^{\prime}(\omega,\tau)$ is compared with the optical absorption spectrum $\alpha(\omega)=(-1/\pi){\rm Im}\left<0\right|j(\omega+E_{0}-H+i\gamma)^{-1}j^{\dagger}\left|0\right>$. We also calculate $O(\tau)=\left<\psi(\tau)\right|\hat{O}\left|\psi(\tau)\right>$ for a static variable $\hat{O}$.

We apply the finte-system DMRG algorithm to calculations of $\alpha^{\prime}(\omega,\tau)$ and $O(\tau)$~\cite{White}. An open boundary condition is taken for the odd number of lattice sites $L$ in order to remove finite-size effects~\cite{Hotta}. Here, we take $L=13$ and the electron number $N_{\rm el}=7$. The density matrix of the system block is defined by $\rho_{i,j}=\sum_{k,\alpha}p^{\alpha}\psi_{i,k}^{\alpha\ast}\psi_{j,k}^{\alpha}$, where $\sum_{\alpha}p^{\alpha}=1$ and $k$ represents a state in the environment block. A set of $\psi^{\alpha}$ is taken to be $\{ \left|0\right>, j^{\dagger}\left|0\right>, \left|\psi(\tau)\right>, j^{\dagger}\left|\psi(\tau)\right>, (\omega+E_{1}-H+i\gamma)^{-1}j^{\dagger}\left|\psi(\tau)\right> \}$ for a calculation of $\alpha^{\prime}(\omega,\tau)$, and $\{ \left|0\right>, j^{\dagger}\left|0\right>, \left|\psi(\tau)\right> \}$ for $O(\tau)$~\cite{Hallberg,Kuhner,Jeckelmann,Matsueda,Cazalilla,Feiguin}. The truncation number is $m=256$. A correction vector $(\omega+E_{1}-H+i\gamma)^{-1}j^{\dagger}\left|\psi(\tau)\right>$ is evaluated by the modified conjugate-gradient method. The time-evolution operator $e^{-iH\tau}$ is devided into $n$ time slices $e^{-iH\tau/n}$, and the fourth-order Runge-Kutta method is applied for each slice. A value of $n$ is determined so that $t\tau/n=0.01$ corresponding to $E_{1}\tau/n <<1$.

Let us first mention the ground state. The model parameters are taken to be $U=10t$, $V=4t$, and $J_{H}=2t$ being appropriate for the manganese oxides. The COI (FM) phase appears in a region that $J_{s}\gtsim 0.12t$ ($J_{s}\ltsim 0.12t$)~\cite{Garcia}. A value of $\left<0\right|\vec{S}_{\rm tot}\cdot\vec{S}_{\rm tot}\left|0\right>$ [$=S_{\rm tot}(S_{\rm tot}+1)$] is minimum (maximum) in the COI (FM) phase. In the COI phase, spatial alternation of the charge density occurs, and the spin structure factor $S(q)$ has a cusp at $q\sim\pi$. In the following calculation, we take $J_{s}=0.2t$, and the Hilbert space is restricted to $S_{\rm tot}^{z}=0$.

\begin{figure}[htbp]
\begin{center}
\includegraphics[width=11cm]{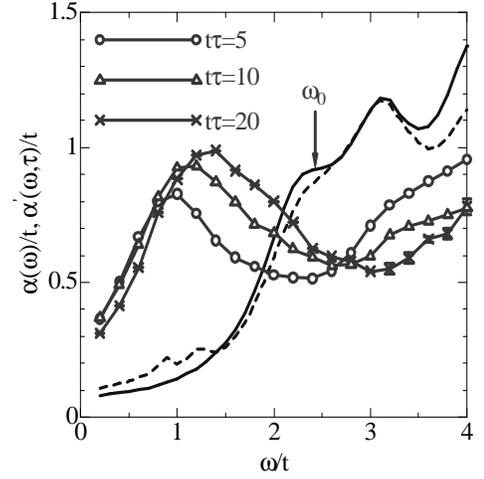}
\end{center}
\caption{$\alpha^{\prime}(\omega,\tau)$ for various $\tau$ values. The solid lines with and without the symbols represent $\alpha^{\prime}(\omega,\tau)$ and $\alpha(\omega)$, respectively. The dashed line represents $\beta(\omega)$.}
\end{figure}

Figure 1 shows $\alpha^{\prime}(\omega,\tau)$ for $t\tau=5$ $(=\gamma_{0}^{-1}+\gamma^{-1})$, $10$ and $20$, and $\alpha(\omega)$~\cite{error}. We take $\omega_{0}=2.4t$ and $\gamma_{0}=\gamma_{1}=0.4t$~\cite{time}. The optical absorption spectrum $\alpha(\omega)$ has a finite charge gap being about $2t$, indicating the COI phase. In contrast to $\alpha(\omega)$, $\alpha^{\prime}(\omega,t\tau=5)$ has a pronounced peak inside the gap at $\omega\sim t$. We have confirmed that this peak shifts toward lower-energy region with increasing the system size $L$. Thus, the system just after the photoirradiation seems to be metallic. With increasing $\tau$, the peak shifts to higher-energy region and broadens. This shift suggests that the system tends to recover the initial COI state, and a part of the kinetic energy of the conduction electrons flows into the localized spins. All of $\alpha^{\prime}(\omega,\tau)$ cross at $\omega\sim 2.8t$, and for $\omega\gtsim 2.8t$, the spectral weight decreases with increasing $\tau$. We have confirmed that there is not remarkable difference between $\alpha^{\prime}(\omega,t\tau=40)$ and $\alpha^{\prime}(\omega,t\tau=20)$.

In order to understand origin of the peak inside the COI gap, we calculate $\beta(\omega)$ defined by $\beta(\omega)=(-1/\pi){\rm Im}\left<0\right|\eta(\omega+E_{0}-H+i\gamma)^{-1}\eta^{\dagger}\left|0\right>$ with $\eta=\eta_{0}-\left<0\right|\eta_{0}\left|0\right>$ and $\eta_{0}=-t\sum_{i,\sigma}(c_{i,\sigma}^{\dagger}c_{i+1,\sigma}+{\rm H.c.})$. This represents distribution of photoexcited states whose parity is different from that of $\alpha(\omega)$. As shown in Fig. 1, the spectra $\alpha(\omega)$ and $\beta(\omega)$ are nearly degenerate for $\omega\ltsim 3.5t$, and $\beta(\omega)$ has a peak at $\omega=3.1t$. The excitation energy of the peak inside the COI gap in $\alpha^{\prime}(\omega,t\tau=5)$ is assigned to a difference between $\omega_{0}$ ($=2.4t$) and $3.1t$, since the initial and final states of transition represented by $\alpha^{\prime}(\omega,\tau)$ have the same parity as $\alpha(\omega)$ and $\beta(\omega)$ have, respectively.

\begin{figure}[htbp]
\begin{center}
\includegraphics[width=11cm]{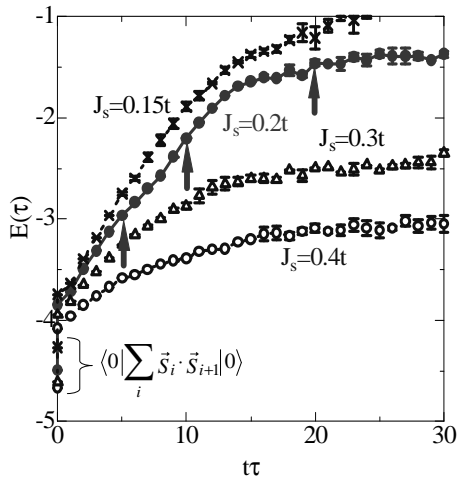}
\end{center}
\caption{$E(\tau)$ and $\left<0\right|\sum_{i}\vec{S}_{i}\cdot\vec{S}_{i+1}\left|0\right>$ for various $J_{s}$ values. For all $J_{s}$ values, we take $\omega_{0}=2.4t$ for which $\alpha(\omega_{0})$ has finite spectral weight. The arrows indicate $t\tau$ values for which the transient absorption spectra are calculated in Fig. 1.}
\end{figure}

In order to examine time evolution of the localized spins, we show $E(\tau)=\left<\psi(\tau)\right|\sum_{i}\vec{S}_{i}\cdot\vec{S}_{i+1}\left|\psi(\tau)\right>$ for various $J_{s}$ values in Fig. 2. We find that $E(\tau)$ increses with $\tau$ and is saturated. Thus, the AF spin alignment in the ground state is disturbed by the photoirradiation. There are two characteristic time scales in $E(\tau)$: (1) the dulation time of the pump photon, $\gamma_{0}^{-1}$, in which a difference between $E(0)$ and $\left<0\right|\sum_{i}\vec{S}_{i}\cdot\vec{S}_{i+1}\left|0\right>$ appears, and (2) relaxation time of the localized spins that is about $t\tau=15$ for $J_{s}=0.2t$. This value of the relaxation time is consistent with that estimated by the $\tau$-dependence of $\alpha^{\prime}(\omega,\tau)$ as mentioned in the previous paragraph. Thus, the charge and spin dynamics are highly correlated with each other. When $J_{s}$ decreases and the system approaches the phase boundary between the COI and FM phases, the $\tau$-dependence of $E(\tau)$ becomes remarkable. This means that the localized spins easily rearrange near the phase boundary.

\begin{figure}[htbp]
\begin{center}
\includegraphics[width=11cm]{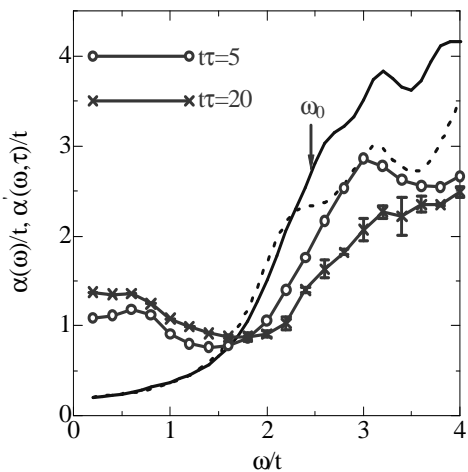}
\end{center}
\caption{$\alpha(\omega)$ and $\alpha^{\prime}(\omega,\tau)$ for the extended Hubbard model. The solid lines with and without the symbols represent $\alpha^{\prime}(\omega,\tau)$ and $\alpha(\omega)$, respectively. The dotted line represents $33\alpha(\omega)/13$ for the extended DE model.}
\end{figure}

Here, we make a comparison of $\alpha^{\prime}(\omega,\tau)$ between the extended DE and Hubbard models in order to illustrate roles of the localized spins on $\alpha^{\prime}(\omega,\tau)$. The 1D extended Hubbard model is defined by
\begin{eqnarray}
H&=&-t\sum_{i,\sigma}( c^{\dagger}_{i,\sigma}c_{i+1,\sigma}+{\rm H.c.} ) \nonumber \\
&& +U\sum_{i}n_{i,\uparrow}n_{i,\downarrow}+V\sum_{i}n_{i}n_{i+1}.
\end{eqnarray}
We take $U=10t$ and $V=6t$ in order to reproduce magnitude of the COI gap in the extended DE model. The other parameters are taken to be $L=33$, $N_{\rm el}=17$ (9 electrons with up spin), and $\omega_{0}=2.4t$. Figure 3 shows $\alpha^{\prime}(\omega,\tau)$ and $\alpha(\omega)$ in the extended Hubbard model. We find that the COI gap collapses immediately after the photoirradiation. In contrast to the extended DE model, the $\tau$-dependence of $\alpha^{\prime}(\omega,\tau)$ is not remarkable, and the recovery of the initial COI phase is not observed for $t\tau\le 20$. These differences of $\alpha^{\prime}(\omega,\tau)$ between the two models come from different characters of their spin correlations. In the extended Hubbard model, AF correlations are weak at quarter filling. On the other hand, in the extended DE model, the AF correlation between the localized spins is strong in the COI phase, and is remarkably influenced by the photoirradiation. This change in the spin correlation causes the $\tau$-dependence of $\alpha^{\prime}(\omega,\tau)$.

We should stress the correlation of the spin and charge dynamics particulary in the extended DE model shown in Figs. 1 and 2. This correlation suggests that change in the charge states due to the photoirradiation strongly affects the spin states. This character of the transient states is in sharp contrast to the spin-charge separation for low-lying states in 1D. This is because high-energy states with various charge and spin configurations are concerned in the early stages of the PIPT.

Now, we discuss implications of the present DMRG results in light of the recent pump-probe spectroscopy measurements on the manganese oxides~\cite{Ogasawara,Miyasaka,Matsubara}. As already shown in Fig. 1, $\alpha^{\prime}(\omega,\tau)$ in the extended DE model has the peak inside the COI gap just after the photoirradiation. This peak shifts to higher-energy region with increasing $\tau$. These results reproduce the pump-probe reflection spectroscopy data~\cite{Ogasawara}. It is also shown that this $\tau$-dependence of $\alpha^{\prime}(\omega,\tau)$ is accompanied by remarkable suppression of the spin structure factor $S(q\sim\pi)$ and slight enhancement of $S(q\sim 0)$. Here, a finite value of magnetization does not appear, since the DMRG calculation has been performed on conditions $S_{\rm tot}^{z}=0$ (the minimum $S_{\rm tot}$) and $[j,\vec{S}_{\rm tot}]=\vec{0}$. These conditions may be reasonable in the manganese oxides where matrix elements of the spin-orbit coupling for the $e_{g}$-orbital vanishes, but seem to be inconsistent with the Kerr rotation data~\cite{Miyasaka,ls}. For the electronic states in the manganese oxides, we suppose that short-range spiral-spin states or microscopic FM domains with keeping $S_{\rm tot}^{z}=0$ are created just after the photoirradiation. Photocarriers can move inside these structures, while magnetization is not detected by the Kerr spectroscopy measurements whose spot size is of the order of 100 $\mu$m $\times$ 100 $\mu$m. When the domain grows into macroscopic one, it is possible that volumes of up- and down-spin domains in the spot are different, and a finite Kerr rotation appears. The above scenario is consistent with recent experiments on Nd${}_{0.5}$Sr${}_{0.5}$MnO${}_{3}$ and Gd${}_{0.55}$Sr${}_{0.45}$MnO${}_{3}$: a low-energy (high-energy) part of the reflectivity increases (decreases) immediately after the photoirradiation, while the Kerr rotation is delayed by the order of 1 ps~\cite{Miyasaka,Matsubara}. The present DMRG data are expected to be applicable to the manganese oxides within a time scale before the Kerr rotation appears.

\begin{figure}[htbp]
\begin{center}
\includegraphics[width=11cm]{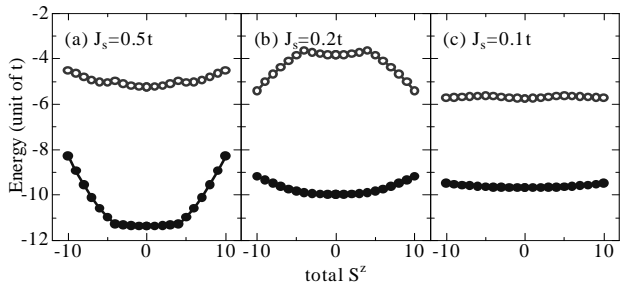}
\end{center}
\caption{Energy-surface diagrams. The filled and open circles represent $E_{0}(S_{\rm tot}^{z})$ and $E_{1}(S_{\rm tot}^{z})$, respectively.}
\end{figure}

Finally, we examine stability of the macroscopic FM domain. Figure 4 shows an energy surface $E_{1}(S_{\rm tot}^{z})=\left<S_{\rm tot}^{z}\right|jHj^{\dagger}\left|S_{\rm tot}^{z}\right>/\left<S_{\rm tot}^{z}\right|jj^{\dagger}\left|S_{\rm tot}^{z}\right>$ as a function of $S_{\rm tot}^{z}$ that represents the energy of the photoexcited state just after the photoirradiation. The lowest-energy eigenstate of $H$ with involved $S_{\rm tot}^{z}$ is represented by $\left|S_{\rm tot}^{z}\right>$. For comparison, we plot $E_{0}(S_{\rm tot}^{z})=\left<S_{\rm tot}^{z}\right|H\left|S_{\rm tot}^{z}\right>$. As shown before, $E_{0}(S_{\rm tot}^{z}=0)$ takes minimum in the COI phase ($J_{s}\gtsim 0.12t$), while $E_{1}(S_{\rm tot}^{z})$ is stable for the maximum $S_{\rm tot}^{z}$ near the phase boundary between the FM and COI phases ($0.12t\ltsim J_{s}<0.5t$). This implies stability of the macroscopic domain in the photoexcited states. In the present scenario, the system after the photoirradiation is located on a point $E_{1}(S_{\rm tot}^{z}=0)$ in Fig. 4(b). Here, the reflectivity shows metallic behavior, while the Kerr rotation does not appear. Then, the system goes to a point $E_{1}(S_{\rm tot}^{z}=10)$. The macroscopic FM domain is created, and the Kerr rotation appears.

In summary, we have examined the effect of the photoirradiation on the COI phase in the 1D extended DE model. We have applied the DMRG method to calculation of the transient absorption spectra. The COI state becomes metallic just after the photoirradiation, and the system tends to go back to the initial COI. This change is accompanied by remarkable suppresion of the AF correlation. There are striking differences of the transient spectra between the extended DE and Hubbard models. The DMRG results are consistent with the recent pump-probe spectroscopy data on the manganese oxides.

We would like to thank K. Miyano, Y. Tokura, H. Okamoto, Y. Okimoto, and M. Matsubara for showing experimental data prior to publication. We also thank A. Takahashi, S. Iwai, N. Shibata, and H. Yokoyama for enlightening discussions. This work was supported by Grant-in-Aid for Scientific Research, JSPS KAKENHI (16340094, 16104005) and TOKUTEI "High Field Spin Science in 100T" (No.451, 18044001) from MEXT, NAREGI, and CREST. H. M. acknowledges the financial support of Department of Physics, Tohoku University.


\begin{thebibliography}{99}
\bibitem{error}
For $m=256$ and $L=13$, we calculate $\alpha^{\prime}(\omega,\tau)$ in two different superblocks where system and environment blocks have  6 and 7 (7 and 6) sites, respectively. The error bars in Figs. 1 and 2 correspond to difference between these data.
\bibitem{time}
This value of $\gamma_{0}^{-1}$ is much smaller the dulation time of the pump photon in the spectroscopy measurements. However, we take this value in order that energy width of the pump photon is larger than energy-level difference in this finite-size cluster.
\bibitem{ls}
Low-symmetry distortion in MnO${}_{6}$ octahedra or GdFeO${}_{3}$-type distortion mixes the $e_{g}$-orbitals with the $t_{2g}$-ones, and may induce finite values of the matrix elements of the spin-orbit coupling.
\end{thebibliography}
\end{document}